\documentclass[twocolumn,myaps,superscriptaddress,myrmp,nobalancelastpage]{myrevtex4}
\usepackage{graphicx,bm}
\usepackage{color}

\def\figurename{\sffamily\selectfont Figure}

\begin{document}

\normalfont\sffamily

\title{Anomalous critical fields in quantum critical superconductors}

\author{C. Putzke}
\affiliation{H. H. Wills Physics Laboratory, University of Bristol, Tyndall Avenue, Bristol, BS8 1TL, United Kingdom.}
\author{P. Walmsley}
\affiliation{H. H. Wills Physics Laboratory, University of Bristol, Tyndall Avenue, Bristol, BS8 1TL, United Kingdom.}
\author{J.D. Fletcher}
\affiliation{National Physical Laboratory, Hampton Road, Teddington, Middlesex TW11 0LW, United Kingdom.}
\author{L. Malone}
\affiliation{H. H. Wills Physics Laboratory, University of Bristol, Tyndall Avenue, Bristol, BS8 1TL, United Kingdom.}
\author{D. Vignolles}
\affiliation{Laboratoire National des Champs Magn\'{e}tiques Intenses (CNRS-INSA-UJF-UPS), 31400 Toulouse, France.}
\author{C. Proust}
\affiliation{Laboratoire National des Champs Magn\'{e}tiques Intenses (CNRS-INSA-UJF-UPS), 31400 Toulouse, France.}
\author{S. Badoux}
\affiliation{Laboratoire National des Champs Magn\'{e}tiques Intenses (CNRS-INSA-UJF-UPS), 31400 Toulouse, France.}
\author{P. See}
\affiliation{National Physical Laboratory, Hampton Road, Teddington, Middlesex TW11 0LW, United Kingdom.}
\author{H.E. Beere}
\affiliation{Cavendish Laboratory, University of Cambridge, J.J. Thomson Avenue, Cambridge CB3 0HE, United Kingdom.}
\author{D.A. Ritchie}
\affiliation{Cavendish Laboratory, University of Cambridge, J.J. Thomson Avenue, Cambridge CB3 0HE, United Kingdom.}
\author{S. Kasahara}
\affiliation{Department of Physics, Kyoto University, Sakyo-ku, Kyoto 606-8502, Japan.}
\author{Y. Mizukami}
\affiliation{Department of Physics, Kyoto University, Sakyo-ku, Kyoto 606-8502, Japan.}
\affiliation{Department of
Advanced Materials Science, University of Tokyo, Kashiwa, Chiba 277-8561, Japan}
\author{T. Shibauchi}
\affiliation{Department of Physics, Kyoto University, Sakyo-ku, Kyoto 606-8502, Japan.} \affiliation{Department of
Advanced Materials Science, University of Tokyo, Kashiwa, Chiba 277-8561, Japan}
\author{Y. Matsuda}
\affiliation{Department of Physics, Kyoto University, Sakyo-ku, Kyoto 606-8502, Japan.}
\author{A. Carrington*}
\affiliation{H. H. Wills Physics Laboratory, University of Bristol, Tyndall Avenue, Bristol, BS8 1TL, United Kingdom.}

\begin{abstract}Fluctuations around an antiferromagnetic quantum critical point (QCP) are believed to lead to
unconventional superconductivity and in some cases to high-temperature superconductivity.  However, the exact mechanism
by which this occurs remains poorly understood. The iron-pnictide superconductor BaFe$_2$(As$_{1-x}$P$_x$)$_2$ is
perhaps the clearest example to date of a high temperature quantum critical superconductor, and so it is a particularly
suitable system in which to study how the quantum critical fluctuations affect the superconducting state. Here we show
that the proximity of the QCP yields unexpected anomalies in the superconducting critical fields. We find that both the
lower and upper critical fields do not follow the behaviour, predicted by conventional theory, resulting from the
observed mass enhancement near the QCP. Our results imply that the energy of superconducting vortices is enhanced,
possibly due to a microscopic mixing of antiferromagnetism and superconductivity, suggesting that a highly unusual
vortex state is realised in quantum critical superconductors.
\end{abstract}

\maketitle
\newpage

\noindent Quantum critical points (QCPs) can be associated with a variety of different order-disorder phenomena,
however, so far superconductivity has only been found close to magnetic order. Superconductivity in heavy fermions,
iron-pnictides, and organic salts is found in close proximity to antiferromagnetic order \cite{Taillefer2010,
Sachdev2011}, whereas in the cuprates the nature of the order (known as the pseudogap phase) is less clear
\cite{Broun2008}. The normal state of these materials have been widely studied and close to their QCPs non-Fermi liquid
behaviour of transport and thermodynamic properties are often found, however, comparatively little is known about how
the quantum critical fluctuations affect the superconducting state \cite{Shibauchi2014}.  This is important as it is
the difference in energy between the normal and superconducting state which ultimately determines the critical
temperature $T_{\rm c}$.

Amongst the various iron-pnictide superconductors, BaFe$_2$(As$_{1-x}$P$_x$)$_2$ has proved to be the most suitable
family for studying the influence of quantum criticality on the superconducting state. This is because the substitution
of As by P introduces minimal disorder as it tunes the material across the phase diagram from a spin-density-wave
antiferromagnetic metal, through the superconducting phase to a paramagnetic metal \cite{Shishido2010}.  The main
effect is a compression of the $c$-axis arising from the smaller size of the P ion compared to As which mimics the
effect of external pressure \cite{Klintberg2010}.  Normal state properties such as the temperature dependence of the
resistivity \cite{Kasahara2010} and spin-lattice relaxation rate \cite{Nakai2010} clearly point to a QCP at $x=0.30$.
Measurements of superconducting state properties that show signatures of quantum critical effects include the magnetic
penetration depth $\lambda$ and the heat capacity jump at $T_{\rm c}$, $\Delta C$ \cite{Hashimoto2012, Walmsley2013}.
Both of these quantities show a strong increase as $x$ tends to 0.30, and it was shown that this could be explained by
an underlying $\sim$6 fold increase in the quasiparticle effective mass $m^*$ at the QCP \cite{Walmsley2013}.

\begin{figure*}
\includegraphics*[width=0.6\linewidth,clip]{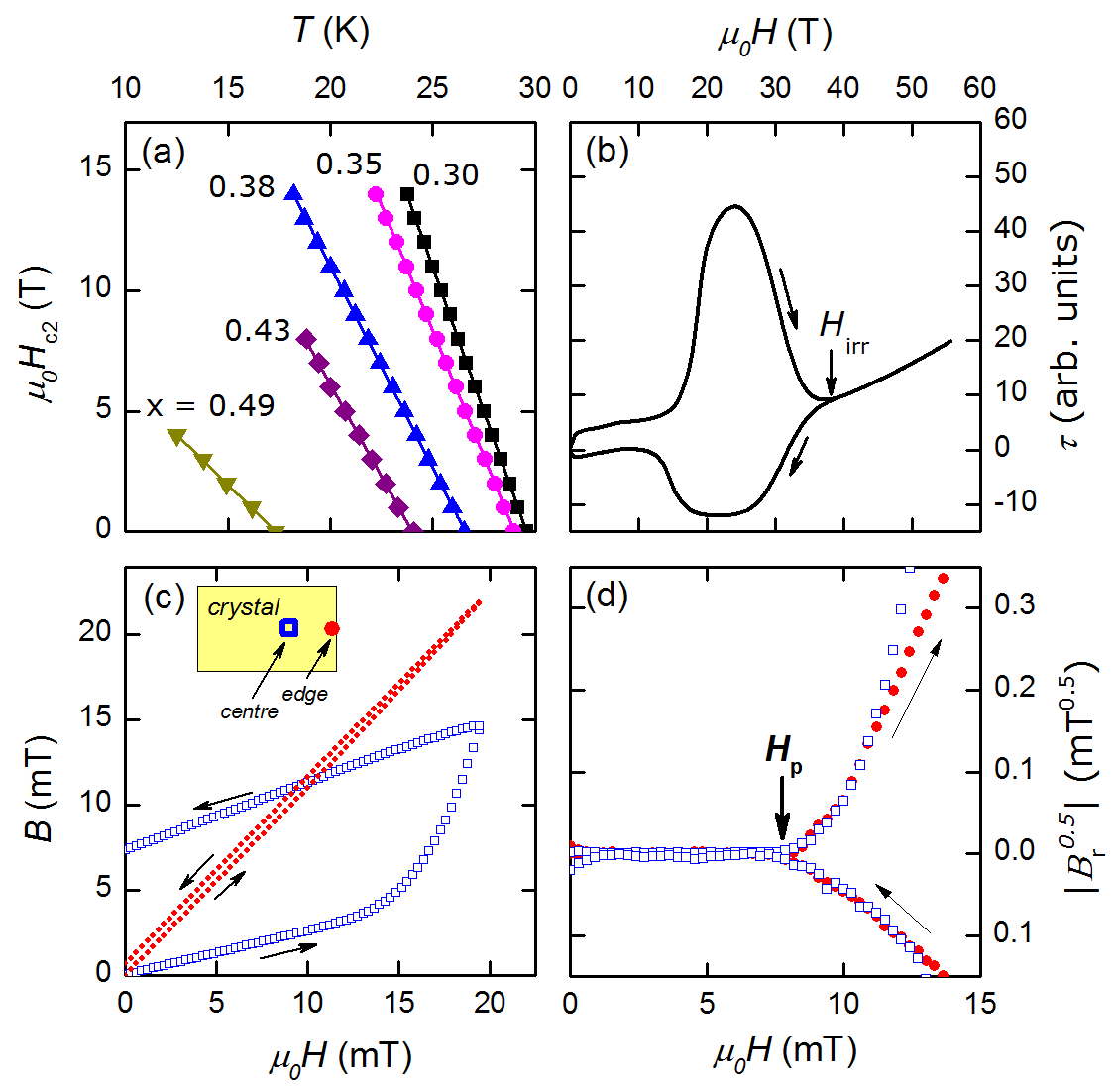}
\caption{{\bf Determination of critical fields.} ({\bf a}), $H_{\rm c2}(T)$ data close to $T_{\rm c}(H=0)$ from heat capacity
measurements for different samples of BaFe$_2$(As$_{1-x}$P$_x$)$_2$. ({\bf b}), Magnetic torque versus rising and falling
field for a sample with $x=0.40$ at $T=1.5$\,K. The irreversibility field $H_{\rm irr}$ is marked. ({\bf c}), Magnetic
flux density $B$ versus applied field $H$ as measured by the micro Hall sensors, for $x=0.35$ and $T=18\,$K at two
different sensor positions: one at the edge of the sample and the other close to the centre (schematic inset). ({\bf d}),
Remnant field $B_{\rm r}$ after subtraction of the linear term due to flux leakage around the sample. $|B_{\rm
r}|^{0.5}$ versus $\mu_0H$ is plotted as this best linearizes  $B_{\rm r}(H)$ \cite{Liang2005}.  Note that the changes
in linearity of $B(H)$ evident in (d) are not visible by eye in (c).}
\end{figure*}

In the standard single-band Ginzburg-Landau theory, the upper critical field is given by
\begin{equation}
H_{\rm c2}=\frac{\phi_0}{2\pi\mu_0\xi_{\rm GL}^2},
\label{eq:hc2}
\end{equation}
where $\phi_0$ is the flux quantum and $\xi_{\rm GL}$ is the Ginzburg-Landau coherence length. In the clean-limit at
low temperature $\xi_{\rm GL}$ is usually well approximated by the BCS coherence length which results in $H_{\rm c2}
\propto (m^*\Delta)^2$, where $m^*$ is the mass of the quasiparticles, and $\Delta$ is the superconducting gap. This
simplified analysis is borne out by the full strong coupling BCS theory \cite{Shulga2002}.  Hence, a strong peak in
$m^*$ at the QCP should result in a corresponding increase in $H_{\rm c2}$ as well as the slope of $H_{\rm c2}$ at
$T_{\rm c}$ ($h^\prime = (dH_{\rm c2}/dT)_{T_{\rm c}}$). This latter quantity is often more easily accessible
experimentally because of the very high $H_{\rm c2}$ values in compounds such as iron-pnictides for $T \ll T_{\rm c}$
and also because the values of $H_{\rm c2}$ close to $T_{\rm c}$ are not reduced by the effect of the magnetic field on
the electron-spin (Pauli limiting effects).

For the lower critical field $H_{\rm c1}$, standard Ginzburg-Landau theory predicts that
\begin{equation}
H_{\rm c1}=\frac{\phi_0}{4\pi \mu_0 \lambda^2}\left(\ln\left(\kappa\right)+0.5\right),
\label{eq:hc1conv}
\end{equation}
where $\kappa=\lambda/\xi_{\rm GL}$, and so the observed large peak in $\lambda$ at the QCP \cite{Hashimoto2012} should
result in a strong suppression of $H_{\rm c1}$.  Here we show that the exact opposite, a peak in $H_{\rm c1}$ at the
QCP, occurs in BaFe$_2$(As$_{1-x}$P$_x$)$_2$, and in addition the expected sharp increase in  $H_{\rm c2}$ is not
observed. This suggest that the critical fields of quantum critical superconductors strongly violate the standard
theory.

\section*{Results}
\noindent\textbf{Upper Critical Field $H_{\rm c2}$}.  We have measured $H_{\rm c2}$ parallel to the $c$-axis, in a
series of high quality single crystal samples of BaFe$_2$(As$_{1-x}$P$_x$)$_2$ spanning the superconducting part of the
phase diagram using two different techniques. Close to $T_{\rm c}(H=0)$ we measured the heat capacity of the sample
using a micro-calorimeter in fields up to 14\,T (see figure 1a). This gives an unambiguous measurement of $H_{\rm
c2}(T)$ and the slope $h^\prime$ which unlike transport measurements is not complicated by contributions from vortex
motion \cite{Serafin2010}.  At lower temperature, we used micro-cantilever torque measurements in pulsed magnetic
fields up to 60\,T. Here, an estimate of $H_{\rm c2}$ was made by observing the field where hysteresis in the torque
magnetisation loop closes (see figure 1b). Although, strictly speaking, this marks the irreversibility line $H_{\rm
irr}$, this is a lower limit for $H_{\rm c2}(0)$ and in superconductors with negligible thermal fluctuations and low
anisotropy such as BaFe$_2$(As$_{1-x}$P$_x$)$_2$ $H_{\rm irr}$ should coincide approximately with $H_{\rm c2}$. Indeed,
in Fig.\,2 we show that the extrapolation of the high temperature specific heat results, using the Helfand-Werthamer
(HW) formula \cite{Helfand1966}, to zero temperature are in good agreement with the irreversibility field measurements
showing both are good estimates of $H_{\rm c2}(0)$.

\begin{figure} \includegraphics*[width=0.95\linewidth,clip]{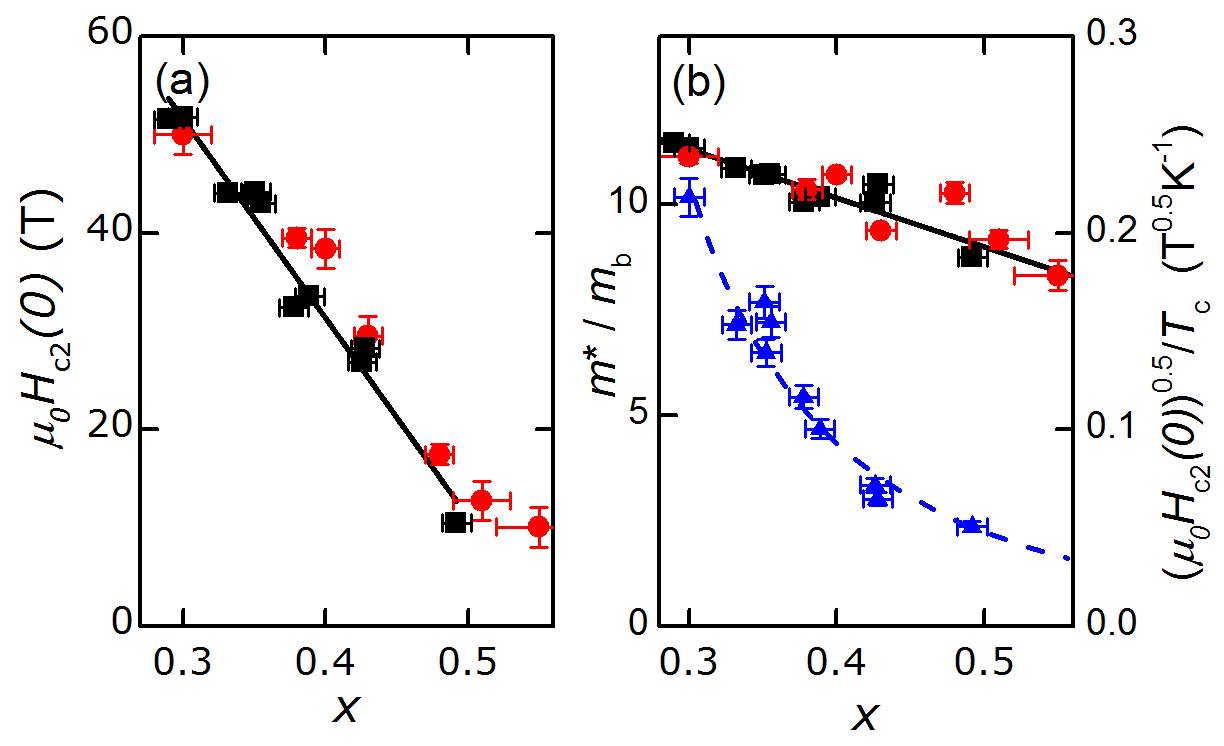} \caption{{\bf Upper critical field as a function
of concentration $x$.} ({\bf a})$H_{\rm c2}(0)$ in BaFe$_2$(As$_{1-x}$P$_x$)$_2$ estimated from the slope of $H_{\rm
c2}(T)$ close to $T_{\rm c}$ using  $H_{\rm c2}(0) = -0.73T_{\rm c} (dH_{\rm c2}/dT)|_{T_{\rm c}}$ (squares)
\cite{Helfand1966}, and also estimates of $H_{\rm c2}(0)$ from the irreversibility field at low temperature
($T=1.5$\,K) measured by torque magnetometry (circles). Error bars on $H_{\rm c2}$ (circles) represent the uncertainties in locating $H_{\rm irr}$ and (squares) in
extrapolating the values close to $T_{\rm c}$ to $T$ = 0. Error bars on $x$ represent standard deviations. ({\bf b}) The same data plotted as $(H_{\rm
c2}(0))^{0.5}/T_{\rm c}$, which, in conventional theory, is proportional to the mass enhancement $m^*$. The mass
renormalization $m^*/m_{\rm b}$ derived from specific heat measurements is shown for comparison (triangles)
\cite{Walmsley2013}. The dashed line is a guide to the eye and solid lines in both parts are linear fits to the data.}
\end{figure}

In the clean-limit we would expect $(H_{\rm c2}(0))^{1/2}/T_{\rm c}$ to be proportional to the renormalized effective
mass $m^*$. Surprisingly, we show in figure 2 that this quantity increases by just $\sim 20$\% from $x=0.47$ to
$x=0.30$ whereas $m^*$ increases by $\sim$400\% for the same range of $x$.

\noindent\textbf{Lower Critical Field $H_{\rm c1}$}.    We measured $H_{\rm c1}$ in our BaFe$_2$(As$_{1-x}$P$_x$)$_2$
samples using a micro-Hall probe array. Here the magnetic flux density $B$ is measured at several discrete points a few
microns from the surface of the sample. Below $H_{\rm c1}$, $B$ increases linearly with the applied field $H$ due to
incomplete shielding of the sensor by the sample. Then, as the applied field passes a certain field $H_{\rm p}$, $B$
increases more rapidly with $H$ indicating that vortices have entered the sample (see figure 1 c,d).   Care must be
taken in identifying $H_{\rm p}$ with $H_{\rm c1}$ because, in some cases, surface pinning and geometrical barriers can
push $H_p$ well above $H_{\rm c1}$. However, in our measurements several different checks, such as the equality of
$H_p$ for increasing and decreasing field \cite{Liang2005}, and the independence of $H_{\rm p}$ on the sensor position
\cite{Okzakai2009}, rule this out (see Methods).

\begin{figure}
\includegraphics*[width=0.95\linewidth,clip]{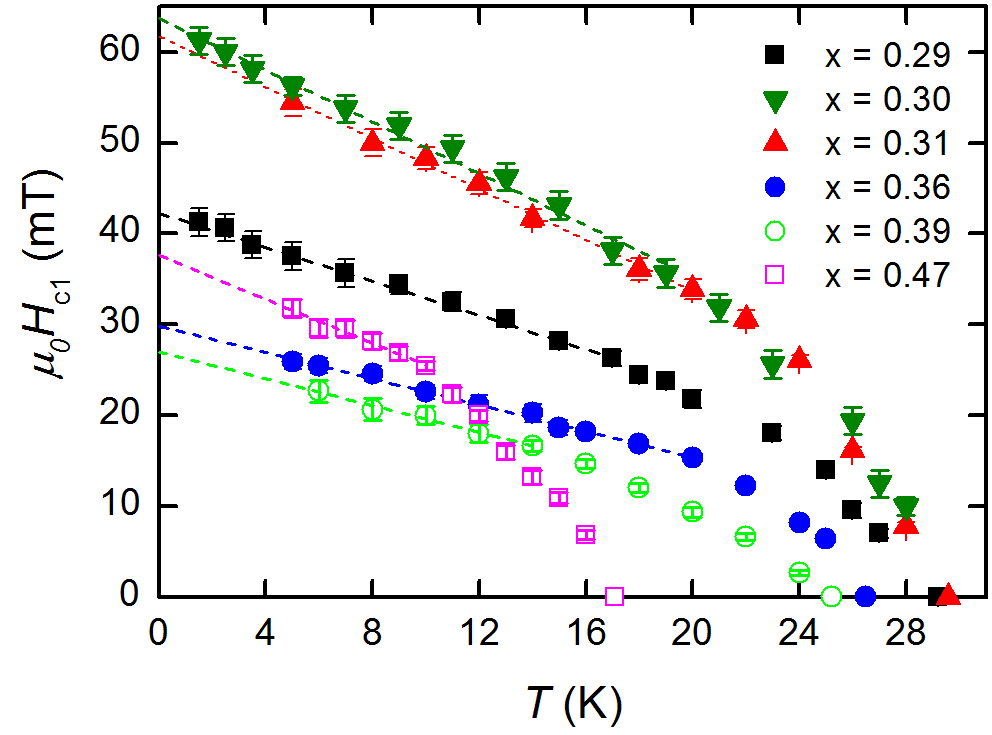}
\caption{{\bf Temperature dependence of $H_{\rm c1}$ in samples of BaFe$_2$(As$_{1-x}$P$_x$)$_2$.} The lines
show the linear extrapolation used to determine the value at $T=0$.
Error bars represent the uncertainty in locating $H_{\rm c1}$ from the raw $M(H)$ data.}
\end{figure}

\begin{figure} \includegraphics*[height=14.5cm,clip]{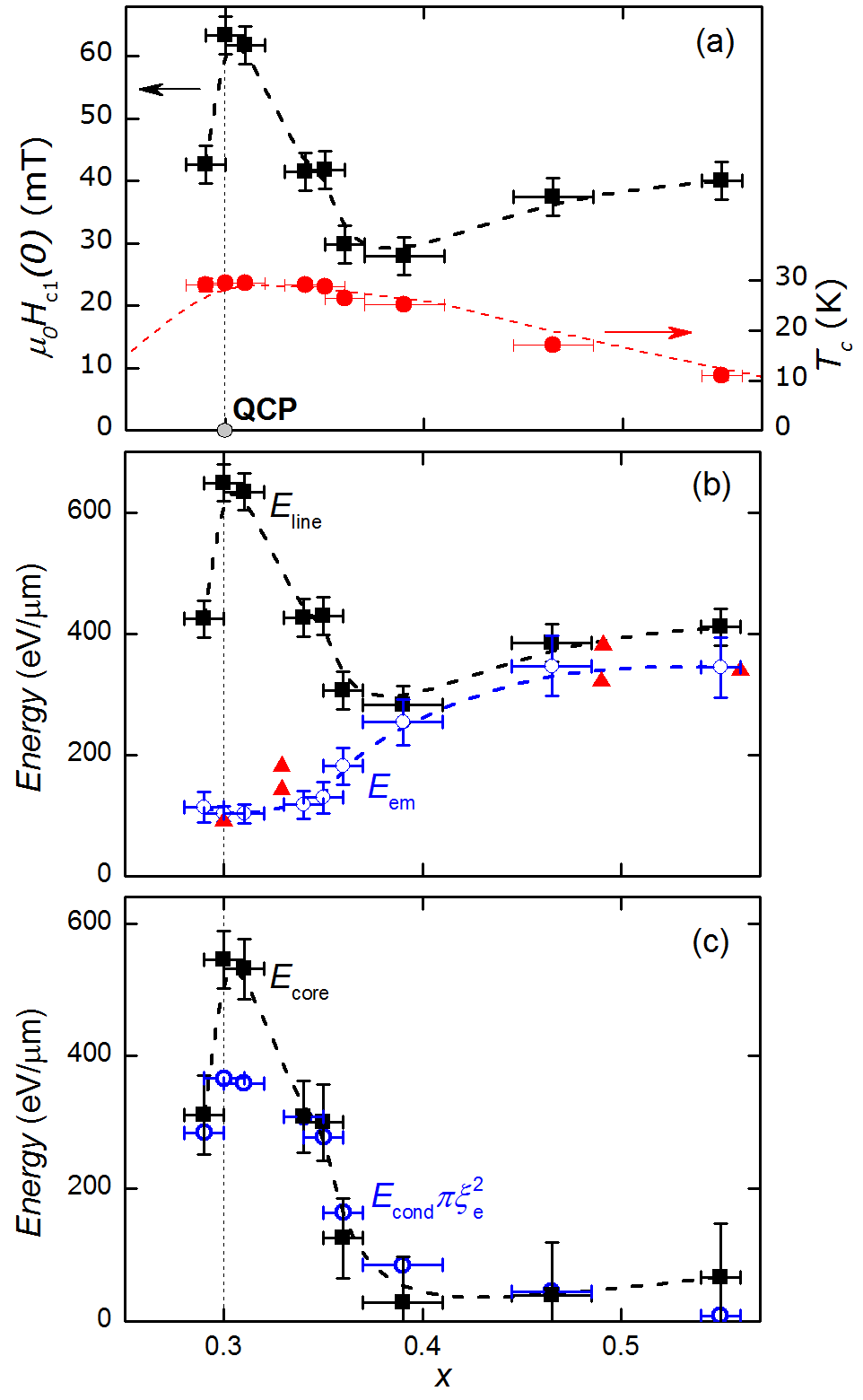} \caption{{\bf Concentration $x$ dependence of
lower critical field and associated energies for BaFe$_2$(As$_{1-x}$P$_x$)$_2$.} ({\bf a}), Lower critical field
$H_{\rm c1}$ extrapolated to $T=0$  and $T_{\rm c}$. The location of the QCP is indicated. Error bars on $H_{\rm c1}$
represent the combination of uncertainties in extrapolating $H_{\rm c1}(T)$ to $T$ = 0 and in the demagnetizing factor.
Error bars on $x$ are standard deviations. ({\bf b}), Vortex line energy $E_{\rm line} =
E_{\rm em}+E_{\rm core}$ at $T=0$ from the $H_{\rm c1}(0)$ data and equations \ref{eq:em} and \ref{eq:hc1gen} shown as
squares. The electromagnetic energy calculated using equation \ref{eq:em} and different estimates of $\lambda$ is also
shown.  The triangles are direct measurements from Ref. \cite{Hashimoto2012}, and the circles are estimates derived by
scaling the band-structure value of $\lambda$ by the effective mass enhancement from specific heat \cite{Walmsley2013}.
Error bars on $E_{\rm em}$ (circles) are calculated from the uncertainty in jump size in heat capacity at $T_{\rm c}$.  ({\bf
c}), Vortex core energy $E_{\rm core} = E_{\rm line}-E_{\rm em}$ along with an alternative estimate derived from the
specific heat condensation energy ($E_{\rm cond}$) and the effective vortex area ($\pi\xi_e^2$).  The uncertainties are calculated from a combination of those in the other panels.
 The dashed lines in all panels are guides to the eye.}
\end{figure}

The temperature dependence of $H_{\rm c1}$ is found to be linear in $T$ at low temperature for all $x$ (figure 3),
which again is indicative of a lack of surface barriers which tend to become stronger at low temperature causing an
upturn in $H_{\rm c1}(T)$ \cite{Burlachkov1992}.   Extrapolating this linear behaviour to zero temperature gives us
$H_{\rm c1}(0)$ which is plotted versus $x$ in Fig.\,4a.   Surprisingly, instead of a dip in $H_{\rm c1}(0)$ at the QCP
predicted by equation \ref{eq:hc1conv} in conjunction with the observed behaviour of $\lambda(x)$ \cite{Hashimoto2012},
there is instead a strong peak.  To resolve this discrepancy we consider again the arguments leading to equation
\ref{eq:hc1conv}.

In general $H_{\rm c1}$ is determined from the vortex line energy $E_{\rm line}$ which is composed of two parts
\cite{Liang94},
\begin{equation}
H_{\rm c1}=(E_{\rm em}+E_{\rm core})/\phi_0.
\label{eq:hc1gen}
\end{equation}
The first,  $E_{\rm em}$ is the electromagnetic energy associated with the magnetic field and the screening currents
which in the high $\kappa$ approximation is given by
\begin{equation}
E_{\rm em} = \frac{\phi_0^2}{4\pi\mu_0\lambda^2} \ln \kappa.
\label{eq:em}
\end{equation}
The second contribution arises from the energy associated with creating the normal vortex core $E_{\rm core}$.   In
high $\kappa$ superconductors, $E_{\rm core}$ is usually almost negligible and is accounted for by the additional
constant 0.5 in equation \ref{eq:hc1conv}.  However, in superconductors close to a QCP we argue this may not be the
case.

In Fig.\,4b,c we use equations \ref{eq:hc1gen} and \ref{eq:em} to determine $E_{\rm em}$ and $E_{\rm core}$. Away from
the QCP, $E_{\rm core}$ is approximately zero and so the standard theory accounts for $H_{\rm c1}(0)$ well. However as
the QCP is approached there is a substantial increase in $E_{\rm core}$ as determined by from the corresponding
increase in $H_{\rm c1}$. We can check this interpretation by making an independent estimate of the core energy from
the condensation energy $E_{\rm cond}$ which we estimate from the experimentally measured specific heat (see Methods).
The core energy is then $E_{\rm cond}\pi\xi^2_{\rm e}$ where $\xi_{\rm e}$ is the effective core radius which may be
estimated from the coherence length $\xi_{\rm GL}$ derived from $H_{\rm c2}$ measurements using Eq.\ \ref{eq:hc2}.  In
Fig.\,4 we see that $E_{\rm cond}\pi\xi^2_{e}$ has a similar dependence on $x$ as $E_{\rm core}$ and is in approximate
quantitative agreement if $\xi_{\rm e} \simeq 4.0\xi_{\rm GL}$ for all $x$. Hence, this suggests that the observed
anomalous increase in $H_{\rm c1}$  could be caused by the high energy needed to create a vortex core close to the QCP.

\section*{Discussion}

\noindent In principle, the relative lack of enhancement in $H_{\rm c2}$ close to the QCP could be caused by impurity
or multiband effects, although we argue that neither are likely explanations.  Impurities  decrease $\xi_{\rm GL}$ and
in the extreme dirty limit $H_{\rm c2}\propto m^*T_{\rm c}/\ell$, where $\ell$ is the electron mean-free-path
\cite{Shulga2002}. Hence, even in this limit we would expect $H_{\rm c2}$ to increase with $m^*$ although not as
strongly as in the clean case.  Impurities increase $H_{\rm c2}$ and as the residual resistance increases close to
$x=0.3$ \cite{Kasahara2010} we would actually expect a larger increase in $H_{\rm c2}$ than expected from clean limit
behaviour. dHvA measurements show that $\ell \gg \xi_{\rm GL}$ at least for the electron bands and for $x>0.38$, which
suggest that, in fact, our samples are closer to the clean limit.

To discuss the effect of multiple Fermi surface sheets on $H_{\rm c2}$ we consider the results of Gurevich
\cite{Gurevich2010} for two ellipsoidal Fermi surface sheets with strong interband pairing.  This limit is probably the
one most appropriate for BaFe$_2$(As$_{1-x}$P$_x$)$_2$ \cite{Hirschfeld2011}.  In this case for $H\|c$, $h^\prime
\propto T_c/(v_1^2+v_2^2)$ were $v_{1,2}$ are the in-plane Fermi velocities on the two sheets. So if the velocity were
strongly renormalized on one sheet only ($v_1\rightarrow 0$) then $H_{\rm c2}$ would be determined mostly by $v_2$ on
the second sheet and hence would not increase with $m^*$ in accordance with our results.  However, in this case the
magnetic penetration depth $\lambda$, which will also be dominated by the Fermi surface sheet with the largest $v$,
would not show a peak at the QCP in disagreement with experiment \cite{Hashimoto2012}. In fact, the numerical agreement
between the increase in $m^*$ with $x$ as determined by $\lambda$ or specific heat, which in contrast to $\lambda$ is
dominated by the low Fermi velocity sections, rather suggests that the renormalization is mostly uniform on all sheets
\cite{Walmsley2013}. In the opposite limit, appropriate to the prototypic multiband superconductor MgB$_2$, where
intraband pairing dominates over interband, $H_{\rm c2}$ will be determined by the band with the lowest $v$
\cite{Gurevich2010} and again an increase in $m^*$ should be reflected in $H_{\rm c2}$. So these multiband effects
cannot easily explain our results.

Another effect of multiband superconductivity is that it can modify the temperature dependence of $H_{\rm c2}$ such
that it departs from the HW model. For example, in some iron-based superconductors a linear dependence of $H_{\rm
c2}(T)$ was found over a wide temperature range \cite{Yeninas2013}.  For BaFe$_2$(As$_{1-x}$P$_x$)$_2$ however, the
coincidence between the HW extrapolation of the  $H_{\rm c2}$ data close to $T_{\rm c}$ and the pulsed field
measurement of $H_{\rm irr}$ for $T\ll T_{\rm c}$ for all $x$, would appear to rule out any significant underestimation
of $H_{\rm c2}(0)$. In Supplementary Figure 3 we show that $H_{\rm irr}$ for a sample with $x=0.51$ fits the HW theory
for $H_{\rm c2}(T)$ over the full temperature range. There is no reason why $H_{\rm irr}$ would underestimate $H_{\rm
c2}(0)$ by the same factor as the HW extrapolation. Even in cuprate superconductors where, unlike here, there is
evidence for strong thermal fluctuation effects, $H_{\rm irr}$ has been shown to agree closely with $H_{\rm c2}$ in the
low temperature limit \cite{Grissonnanche2014}. The magnitude of the discrepancy between the behaviour of $H_{\rm
c2}(0)$ and $m^*$ discussed above (see figure 2) also makes an explanation based on an experimental underestimate of
$H_{\rm c2}(0)$ implausible.

Another possibility is that in heavy fermion superconductors the mass enhancement is often reduced considerably at high
fields and so therefore $m^*$ could be reduced at fields comparable to $H_{\rm c2}$. In BaFe$_2$(As$_{1-x}$P$_x$)$_2$
however, a significantly enhanced mass in fields greater than $H_{\rm c2}$ can be inferred from the dHvA measurements
\cite{Walmsley2013} and low temperature, high field, resistivity \cite{Analytis2014}.  Although very close to the QCP
the mass inferred from these measurements is slightly reduced from the values inferred from the zero field specific
heat measurements \cite{Walmsley2013} this cannot account for the lack of enhancement of $H_{\rm c2}$ shown in figure
2.

Our results are similar to the behaviour observed in another quantum critical superconductor, CeRhIn$_5$. Here the
pressure tuned QCP manifests a large increase in the effective mass as measured by the dHvA effect and the low
temperature resistivity. $T_{\rm c}$ is maximal at the QCP but $H_{\rm c2}$ displays only a broad peak, inconsistent
with the mass enhancement shown by the other probes \cite{Knebel2008}. We should note that in this system $H_{\rm c2}$
at low temperatures is Pauli limited. However,  close to $T_{\rm c}$, $H_{\rm c2}$ is always orbitally limited and as
neither $h^\prime$ or $H_{\rm c2}(0)$ are enhanced in BaFe$_2$(As$_{1-x}$P$_x$)$_2$ or CeRhIn$_5$ \cite{Knebel2008},
Pauli limiting can be ruled out as the explanation.

A comparison to the behaviour observed in cuprates is also interesting.  Here two peaks in $H_{\rm c2}(0)$ as a
function of doping $p$ in YBa$_2$Cu$_3$O$_{7-\delta}$ have been reported \cite{Grissonnanche2014}, which approximately
coincide with critical points where other evidence suggests that the Fermi surface reconstructs.  Quantum oscillation
measurements indicate that $m^*$ increases close to these points \cite{Sebastian2010}, suggesting a direct link between
$H_{\rm c2}(0)$ and $m^*$ in the cuprates in contrast to our finding here for BaFe$_2$(As$_{1-x}$P$_x$)$_2$.  However,
by analysing the data in the same way as we have done here, it can be seen \cite{Tafti2014} that $H_{\rm
c2}(0)^{0.5}/T_{\rm c}$ for YBa$_2$Cu$_3$O$_{7-\delta}$ is independent of $p$ above $p\simeq 0.18$ and falls for $p$
below this value, reaching a minimum at  $p\simeq 1/8$.  This suggest that at least the peak at higher $p$ is driven by
the increasing gap value rather than a peak in $m^*$, in agreement with our results here, and that the minimum in
$H_{\rm c2}(0)^{0.5}/T_{\rm c}$ coincides with the doping where charge order is strongest at $p\simeq 1/8$
\cite{Huecker2014}.

The lack of enhancement of $H_{\rm c2}(0)$ in  all these systems  suggests a fundamental failure of theory.  One
possibility is that this may be driven by microscopic mixing of superconductivity and antiferromagnetism close to the
QCP.  In the vicinity of the QCP, antiferromagnetic order is expected to emerge near the vortex core region where the
superconducting order parameter is suppressed \cite{Demler2004,Zhang2002}.  Such a field-induced antiferromagnetic
order has been observed experimentally in cuprates \cite{Lake2004,Kakuyanagi03}. When the QCP lies beneath the
superconducting dome, as in the case of BaFe$_2$(As$_{1-x}$P$_x$)$_2$ \cite{Hashimoto2012,Shibauchi2014},
antiferromagnetism and superconductivity can coexist on a microscopic level. In such a situation, as pointed out in
Ref.\,\cite{Zhang2002}, the field-induced antiferromagnetism can extend outside the effective vortex core region where
the superconducting order parameter is finite. Such an extended magnetic order is expected to lead to further
suppression of the superconducting order parameter around vortices. This effect will enlarge the vortex core size,
which  in turn will suppress the upper critical field in agreement with our results.  We would expect this effect to be
a general feature of superconductivity close to an antiferromagnetic QCP, but perhaps not relevant to the behaviour
close to $p=0.18$ in the cuprates.

To explain the $H_{\rm c1}$ results we postulate that the vortex core size is around 4 times larger than the estimates
from $H_{\rm c2}$.  This is in fact expected in cases of multiband superconductivity or superconductors with strong gap
anisotropy. In MgB$_2$ \cite{Eskildsen2002,Koshelev2003} and also in the anisotropic gap superconductor
2\textit{H}-NbSe$_2$ \cite{Hartmann1994} the effective core size has been found to be around 3 times $\xi_{\rm GL}$,
similar to that needed to explain the behaviour here. BaFe$_2$(As$_{1-x}$P$_x$)$_2$ is known to have a nodal gap
structure \cite{Hashimoto2010} which remains relatively constant  across the superconducting dome \cite{Hashimoto2012}
and so we should expect the core size to be uniformly enhanced for all $x$. The peak in $H_{\rm c1}(x)$ at the QCP is
then, primarily caused by the fluctuation driven enhancement in the normal state energy, but the effect is magnified by
the nodal gap structure of BaFe$_2$(As$_{1-x}$P$_x$)$_2$.

We expect the observed anomalous increase in $H_{\rm c1}$ to be a general feature of quantum critical superconductors
as these materials often have nodal or strongly anisotropic superconducting gap structures and the increase in normal
state energy is a general property close to a QCP. The relative lack of enhancement in $H_{\rm c2}$ also seems to be a
general feature, which may be linked to a microscopic mixing of antiferromagnetism and superconductivity.

\section*{Methods}
\small \noindent\textbf{Sample growth and characterisation.} BaFe$_2$(As$_{1-x}$P$_x$)$_2$ samples were grown using a
self flux technique as described in Ref.\ \cite{Kasahara2010}. Samples for this study were screened using specific heat
and only samples with superconducting transition width less than 1\,K were measured (see Supplementary Figure 1). To
determine the phosphorous-concentration in the samples we carried out energy-dispersive x-ray analysis (EDX) on several
randomly chosen spots on each crystal ($H_{\rm c1}$ samples) or measured the $c$-axis lattice parameter using x-ray
diffraction ($H_{\rm c2}$ samples) which scales linearly with $x$. For some of the $H_{\rm c2}$ samples measured using
high field torque magnetometry the measured de Haas-van Alphen frequency was also used to determine $x$ as described in
Ref.\ \cite{Walmsley2013}.

\noindent\textbf{Measurements of $H_{\rm c2}$.} Close to $T_{\rm c}$ the upper critical field was determined using heat
capacity. For this a thin film microcalorimeter was used \cite{Walmsley2013}. We measured the superconducting
transition at constant magnetic field up to 14\,T (see Supplementary Figure 2). The midpoint of the increase in $C$ at
the transition defines $T_{\rm c}(H)$.  At low temperatures ($T \ll T_{\rm c}$) we used piezo-resistive
microcantilevers to measure magnetic torque in pulsed magnetic field and hence determine the irreversibility field
$H_{\rm irr}$. The crystals used in the pulsed field study were the same as those used in Ref. \cite{Walmsley2013} for
the de Haas-van Alphen effect (except samples for $x\simeq 0.3$). By taking the difference between the torque in
increasing and decreasing field we determined the point at which the superconducting hysteresis closes as $H_{\rm irr}$
(see figure 1(b)).  For some compositions we measured $H_{\rm irr}$ in dc field over the full temperature range and
found it to agree well with the HW model and also the low temperature measurements in pulsed field on the same sample
(Supplementary figure 3). Our heat capacity measurements of $H_{\rm c2}$ close to $T_{\rm c} (H=0)$ are in good
agreement with those of Ref.\ \cite{Chaparro2012}.

\noindent\textbf{Measurements of $H_{\rm c1}$.} The measurements of the field of first flux penetration $H_{\rm p}$
have been carried out using micro-Hall arrays. The Hall probes were made with either GaAs/AlGaAs heterostructures
(carrier density $n_s=3.5\times 10^{11}\rm{cm}^{-2}$) or GaAs with a 1$\mu$m thick silicon doped layer (concentration
$n_s=1\times 10^{16}\rm{cm}^{-3}$). The latter had slightly lower sensitivity but proved more reliable at temperatures
below 4\,K. The measurements were carried out using a resistive magnet so that the remanent field during zero field
cooling was as low as possible. The samples was warmed above $T_{\rm c}$ after each field sweep and then cooled at a
constant rate to the desired temperature.

When strong surface pinning is present $H_{\rm p}$ may be pushed up significantly beyond $H_{\rm c1}$.  In this case
there will also be a significant difference between the critical field $H_{\rm p}$ measured at the edge and the centre
of the sample (for example see Ref.\ \cite{Okzakai2009}) and also a difference between the field where flux starts to
enter the sample and the field at which it leaves. Some of our samples, also showing signs of inhomogeneity, such as
wide superconducting transitions, showed this behaviour.  An example is shown in supplementary figure 4. In this sample
the sensor at the edge shows first flux penetration at $H_{\rm p}\approx 5$\,mT whereas the value is $\sim 3$ times
higher at the centre.  For decreasing fields, the centre sensor shows a similar value to the edge sensor.   All the
samples reported in this paper showed insignificant difference between $H_{\rm p}$ at the centre and the edge  and also
for increasing and decreasing fields. Hence, we conclude that $H_{\rm c1}$ in our samples is not significantly
increased by pinning.

As our samples are typically thin platelets, demagnetisation effects need to be taken into account for measurement of
$H_{\rm c1}$. Although an exact solution to the demagnetisation problem is only possible for ellipsoids and infinite
slabs, a good approximation for thin slabs has been obtained by Brandt \cite{Brandt1999}. Here $H_{\rm c1}$ is related
to the measured $H_{\rm p}$, determined from $H$ using
\begin{eqnarray}
    H_{\rm c1} = \frac{H_{\rm p}}{\tanh\sqrt{0.36 l_{\rm c}/l_{\rm a}}}
\end{eqnarray}
where $l_{\rm c}$ is the sample dimension along the field and $l_a$ perpendicular to the field.

All samples in this study had $l_{\rm c} \ll l_{\rm a}$. To ensure that the determination of the effective field is
independent of the specific dimension we have carried out multiple measurements on a single sample cleaved to give
multiple ratios of $l_{\rm c}/l_{\rm a}$. The results of this study (supplementary figure 5) show that $H_{\rm c1}$
determined by this method are independent of the aspect ratio of the sample.  Furthermore, the samples used all had
similar $l_{\rm c}/l_{\rm a}$ ratios (see Supplementary Table 1), and so any correction would not give any systematic
errors as a function of $x$.

\noindent\textbf{Calculation of condensation energy.} The condensation energy can be calculated from the specific heat
using the relation
\begin{equation}
E_{\rm cond} = \int_0^\infty \! \left[C_{\rm s}(T)-C_{\rm n}(T)\right] dT.
\label{Eq.Cond}
\end{equation}
To calculate this we first measured a sample of BaFe$_2$(As$_{1-x}$P$_x$)$_2$ with $x=0.47$, using a relaxation
technique in zero field and $\mu_0 H=14$\,T which is sufficient at this doping to completely suppress superconductivity
and thus reach the normal state.  We used this 14\,T data to determine the phonon heat capacity and we then subtract
this from the zero field data to give the electron specific heat of the sample.   We then fitted this data to a
phenomenological nodal gap, alpha model (with variable zero temperature gap) similar to that described in Ref.\
\cite{Taylor2007} (see supplementary figure 6). We then integrated this fit function using Eq.\ \ref{Eq.Cond} to give
$E_{\rm cond}$ for this value of $x$.  For lower values of $x$ (higher $T_{\rm c}$) the available fields were
insufficient to suppress superconductivity over the full range of temperature, so we assumed that the shape of the heat
capacity curve does not change appreciably with $x$ but rather just scales with $T_c$ and the jump height at $T_{\rm
c}$.  This is implicitly assuming that the superconducting gap structure does not change appreciably with $x$, which is
supported by magnetic penetration depth $\lambda$ measurements which show that normalised temperature dependence
$\lambda(T)/\lambda(0)$ is relatively independent of $x$ \cite{Hashimoto2012}.  With this assumption we can then
calculate $$E_{\rm cond}(x) = \frac{E_{\rm cond}(x_{\rm ref}) T_{\rm c}(x) \Delta C(x)}{T_{\rm c}(x_{\rm ref}) \Delta
C(x_{\rm ref})},$$ where $x_{\rm ref}=0.47$.

\def\bibsection{\vskip 6pt \setlength{\parindent}{0pt}{\sffamily\bfseries\selectfont References}  \setlength{\parindent}{12pt}}

\section*{Acknowledgements}

We thank Igor Mazin and Georg Knebel for useful discussions, and A. M. Adamska for experimental help.  This work was
supported by the Engineering and Physical Sciences Research Council (Grant No. EP/H025855/1), EuroMagNET II under the
EU Contract No. 228043, National Physical Laboratory Strategic Research Programme, and KAKENHI from JSPS.

\section*{Author contributions}

A.C. and C.Putzke. conceived the experiment. C.Putzke performed the high field torque measurements (with D.V., C.Proust
and S.B.) and the Hall probe measurements.  P.W. and L.M. performed heat capacity measurements. The Hall probe arrays
were fabricated by J.D.F., P.S., H.E.B and D.A.R.   Samples were grown and characterised by S.K., Y. Mizukami, T.S. and
Y.Matsuda.  The manuscript was written by A.C. with input from  C.Putzke, C. Proust, P.W., L.M., J.D.F, T.S and Y.
Matsuda.

\def\figurename{\sffamily\selectfont\textbf{Supplementary Figure}}
\def\tablename{\sffamily\selectfont\textbf{Supplementary Table}}

\newcommand{\etal}{\emph{et. al.}}
\newcommand{\Hcl}{$H_{c1}$}


\Large\sffamily

\begin{widetext}
\newpage
\textbf{\Large Supplementary information}
\end{widetext}

\begin{figure}[!h]
    \centering
    \includegraphics*[width=0.5\linewidth,clip]{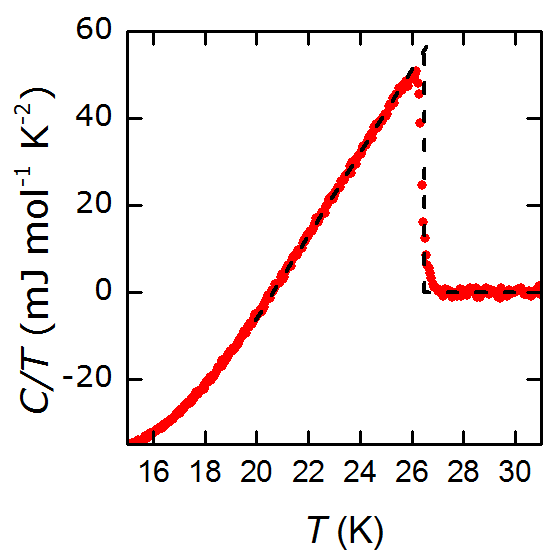}
    \caption{Heat capacity of a sample with $x=0.38$ with the normal state heat capacity subtracted. The dashed line is the behaviour expected for a mean-field like jump.}
    \label{fig:HC}
\end{figure}

\begin{figure}[!h]
    \centering
    \includegraphics*[width=0.7\linewidth,clip]{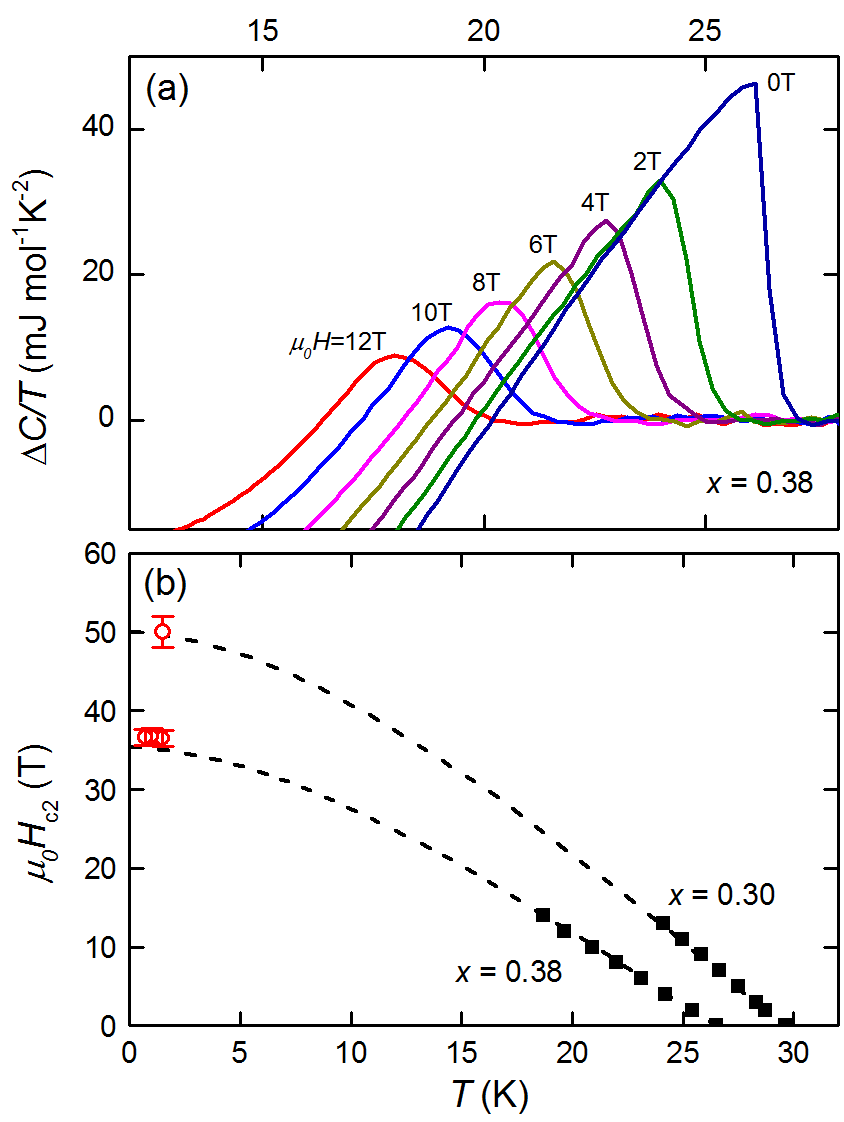}
    \caption{Determination of $H_{\rm c2}$. (a) Heat capacity jump of a sample with $x=0.38$ at various magnetic fields up to 12\,T. A polynomial fit was used to subtract the normal state specific heat.
    Panel (b) shows  $H_{\rm c2}$ versus temperature for the $x=0.38$ sample shown in (a) determined from the midpoint of the jump in the heat capacity and similar data for a sample with $x=0.30$.
    The open circles are values of $H_{\rm irr}$ as determined by torque magnetometry in pulsed field on two other samples with the same values of $x$ and $T_{\rm c}$.
    The dashed lines are fits to the clean limit HW model [14] for $H_{\rm c2}(T)$.}
    \label{fig:HCHc2}
\end{figure}

\begin{figure}[!h]
    \centering
    \includegraphics*[width=0.7\linewidth,clip]{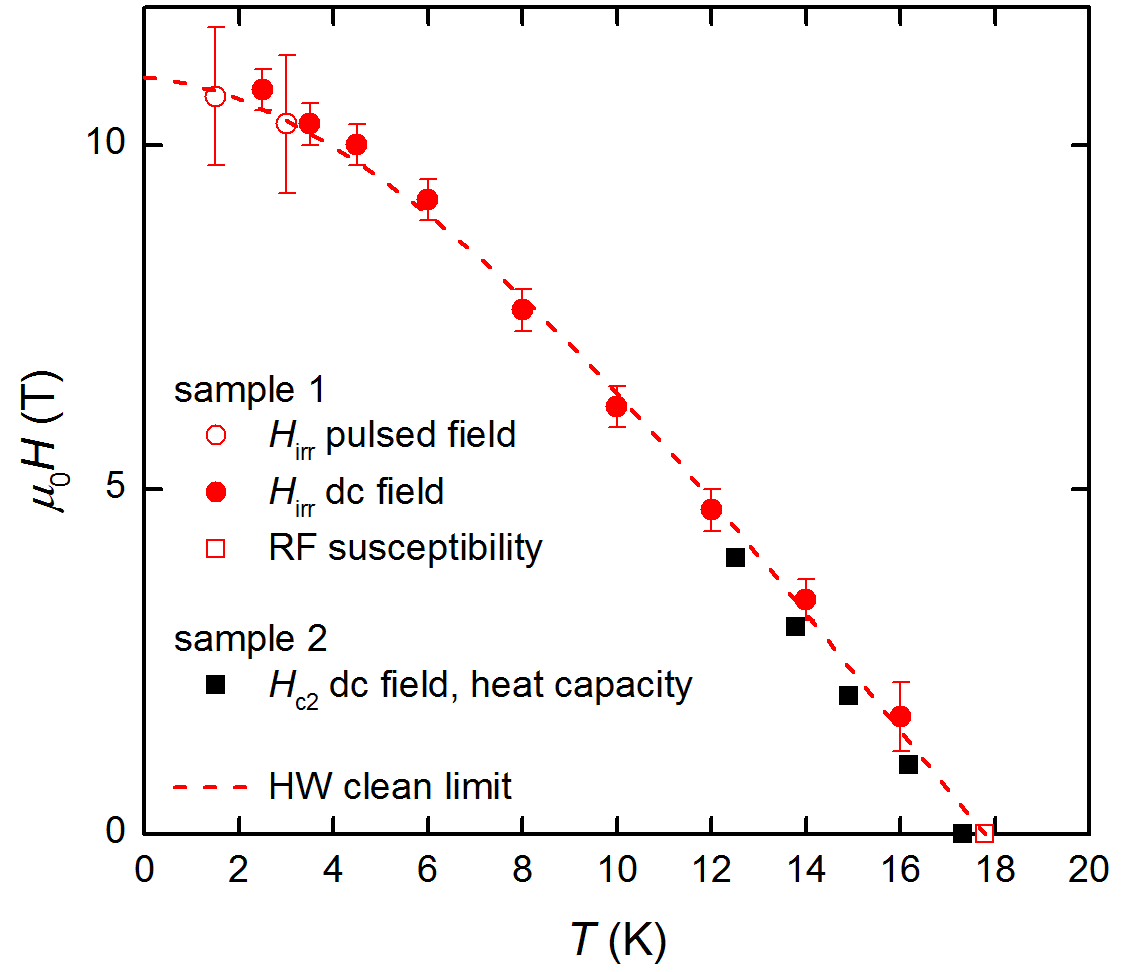}
    \caption{Irreversibility field $H_{\rm irr}$, as determined by torque measurements, versus temperature for a sample with $x=0.51\pm0.02$ over the full temperature range.
    The solid circles are measurements in dc field and the open circles are in pulsed field (same sample). The $T_{\rm c}$ in zero field of this sample was determined
    by very low field radio-frequency susceptibility measurements. The solid squares are measurements of $H_{\rm c2}$ determined from the midpoint of the jump in the specific heat
    of a second sample with almost the same $x$ ($T_{\rm c}$ of this sample is 0.4\,K lower than the torque sample in zero field,
    and $x$ is the same within error by EDX). The dashed line is a fit to the clean limit HW model [14] for $H_{\rm c2}(T)$.}
    \label{fig:HCHirr}
\end{figure}

\begin{figure}[!h]
    \centering
    \includegraphics*[width=0.75\linewidth,clip]{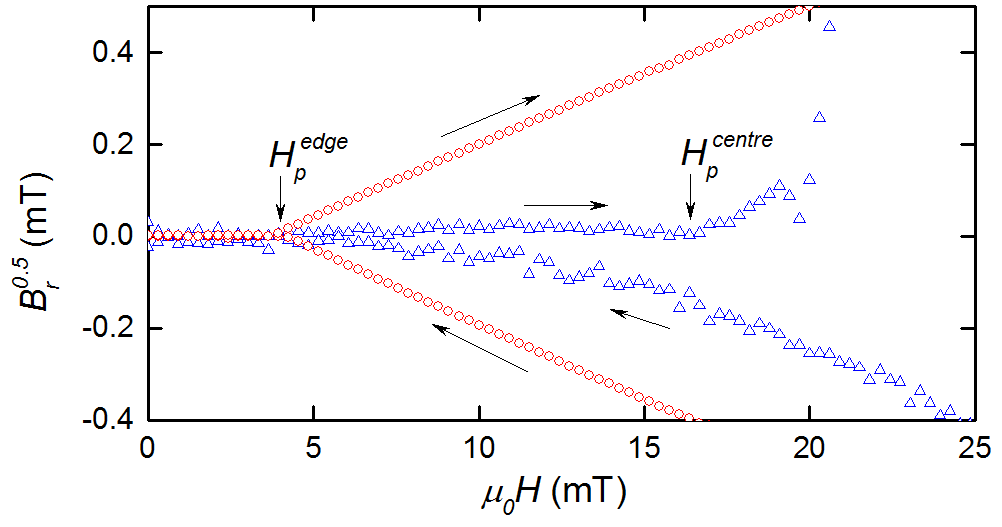}
    \caption{Remnant field $B_{\rm r}$ of a sample with $x=0.30$ that showed a wide superconducting transition of $\Delta T_{\rm c}>1.5\,$K.
    Two sensors at the edge (circles) and at the centre of the sample (triangles) are shown.}
    \label{fig:Hc1fail}
\end{figure}

\begin{figure}[!h]
    \centering
    \includegraphics*[width=0.7\linewidth,clip]{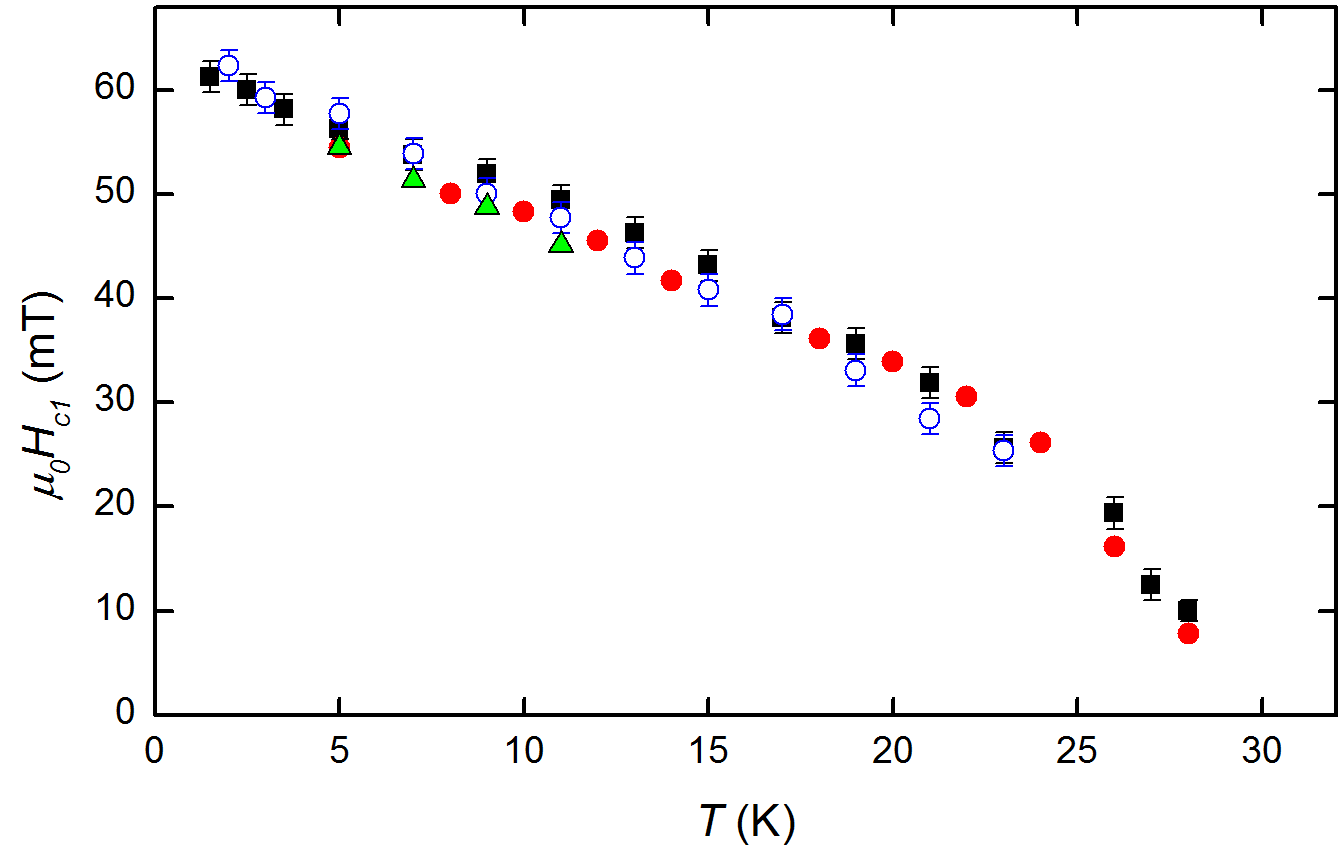}
    \caption{$H_{\rm c1}(T)$ for samples C24 and C2a with $x=0.3$ and $x=0.31$ respectively.
    Sample C24 has been cleaved repeatedly to produce samples with different $l_{\rm c}/l_{\rm a}$ ratios: 0.05 (C24,open circles), 0.08(C24,squares), 0.09 (C2a,filled circles) and 0.11 (C24,triangles), in order
    to test the accuracy of the demagnetising factor determination.}
    \label{fig:Ba122_diffN}
\end{figure}

\begin{figure}[!h]
    \centering
    \includegraphics*[width=0.7\linewidth,clip]{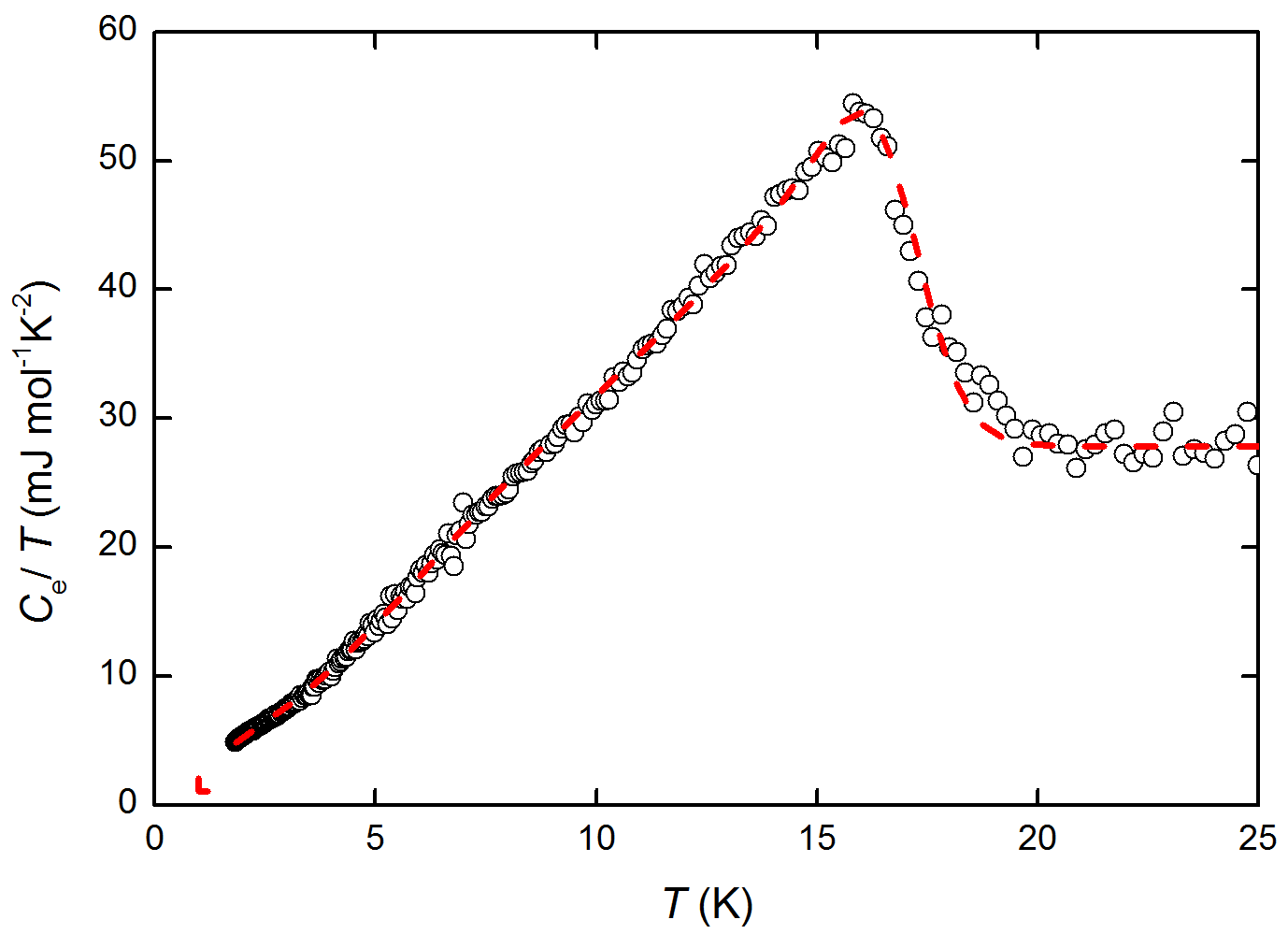}
    \caption{Heat capacity of a $x=0.47$ sample with the phonon contribution subtracted. The phonon contribution was determined directly by using a high field (14,T) to suppress the superconductivity.
    The dashed line is a fit to the data using a nodal gap alpha model similar to Ref.\ 36. The model has been convoluted with a Gaussian distribution to model the spread of $T_{\rm c}$ in the sample.}
    \label{fig:Ba122_AlphaModel}
\end{figure}

\begin{table}[!h]
\renewcommand\thetable{1}
\large
\begin{tabular*}{.6\linewidth}{@{\extracolsep{\fill}}|ccccc|}
\hline
Sample & $x$ & $\Delta x$ & $l_c$ ($\mu$m) & $l_a$ ($\mu$m) \\
\hline
C19 & 0.29 & 0.01 & 11 & 149 \\
C24a & 0.30 & 0.01 & 27 & 360 \\
C24b & 0.30 & 0.01 & 18 & 360 \\
C24c & 0.30 & 0.01 & 18 & 170 \\
C2a & 0.31 & 0.01 & 28 & 300 \\
C21 & 0.34 & 0.01 & 20 & 255 \\
0p3B & 0.35 & 0.01 & 48 & 115 \\
C1 & 0.36 & 0.01 & 35 & 269 \\
C7a & 0.39 & 0.02 & 17 & 260 \\
C9 & 0.47 & 0.02 & 40 & 300\\
0p6a & 0.55 & 0.01 & 48 & 240 \\
\hline
\end{tabular*}
\caption{List of dimension of the samples used for the $H_{\rm c1}$ measurements.} \label{tbl:Dimension}
\end{table}

\end{document}